\documentclass[twoside]{dis08}
\usepackage[latin1]{inputenc}
\usepackage[dvips]{graphicx,epsfig,color}
\usepackage{wrapfig,rotating}
\usepackage{amssymb,amsmath,array}

\newcommand{\etal}{ {\it et al.,} }
\newcommand{\eg} { {\it e.g.} }

\newcommand{\Lumi}{{\cal L}}

\newcommand{\pb}{\,\text{pb}}
\newcommand{\fb}{\,\text{fb}}
\def\z0{Z^0}

\def\as{\alpha_s}
\def\oalphas2{{\cal O}(\alpha\as^2)}

\def\be{\begin{equation}}
\def\ee{\end{equation}}
\def\bea{\begin{eqnarray}}
\def\eea{\end{eqnarray}}

\begin{document}
\title{LHC Startup}

%***********************************************************************
% AUTHORS INFORMATION AREA
%***********************************************************************
\author{Wesley H. Smith
%
% Optional short acknowledgment: remove next line if non-needed
%\thanks{This is an optional funding source acknowledgment.}
%
% DO NOT MODIFY THE FOLLOWING '\vspace' ARGUMENT
\vspace{.3cm}\\
%
% Addresses and institutions (remove "1- " in case of a single institution)
University of Wisconsin Physics Department\\
Madison, WI 53706 USA
%
% Remove the next three lines in case of a single institution
%\vspace{.1cm}\\
%2- School of Second Author - Dept of Second Author \\
%Address of Second Author's school - Country of Second Author's school\\
}
%***********************************************************************
% END OF AUTHORS INFORMATION AREA
%***********************************************************************

\maketitle

\begin{abstract}

The Large Hadron Collider will commence operations in the latter half of 2008. The plans of the LHC experiments ALICE, ATLAS, CMS and LHCb are described. The scenario for progression of luminosity and the strategies of these 4 experiments to use the initial data are detailed. There are significant measurements possible with integrated luminosities of 1, 10 and 100 $\pb^{-1}$. These measurements will provide essential calibration and tests of the detectors, understanding of the Standard Model backgrounds and a first oportunity to look for new physics.

\end{abstract}
\section{Introduction}
\label{sec:intro}

The LHC will eventually operate at 7 times the energy and 100 times the luminosity of the Fermilab Tevatron and will search for the mechanism of particle mass generation, supersymmetry which introduces a partner for each known particle, stabilizes the higgs mass and provides a candidate for the dark matter making up most of the universe, extra dimensions and other exciting new physics possibilities. The LHC will start taking physics data in 2008 with an expected integrated luminosity of less than 100 $pb^{-1}$ at a 10 TeV center of mass energy followed by 14 TeV running in 2009 that is expected to climb to the design annual integrated luminosity of 100 $fb^{-1}$ over the following years.  To prepare for the initial startup period, the LHC experiments have studied simulated data sets corresponding to the expected initial integrated luminosities of 1, 10 and 100 $\pb^{-1}$ at 14 TeV\footnote{The studies cited in this report were performed before the announcement of the 10 TeV LHC startup energy, but are applicable to 14 TeV as noted in Section~\ref{sec:lumi}.}. 

The ATLAS and CMS experiments are general purpose pp detectors. ATLAS uses an inner superconducting solenoid containing silicon tracking and a transition radiation tracker and large superconducting air-core toroids interspersed with muon chambers surrounding liquid argon and iron-scintillator calorimetry. The CMS experiment consists of a large diameter superconducting solenoid containing an all-silicon tracker, a $PbWO_3$ electromagnetic calorimeter and a brass-scintillator hadron calorimeter, with a return yoke and muon chambers outside. The production of bb pairs at the LHC will be peaked in the forward direction, therefore LHCb is designed as a single-arm spectrometer covering 10-300 mrad in polar angle. The LHCb experiment is a single-arm spectrometer consisting of a vertex locator and main tracker, ring imaging Cherenkov counters, an electromagnetic calorimeter, a hadron calorimeter and a muon detector.  The ALICE experiment is designed for heavy ion collisions and consists of a silicon inner tracking system, a cylindrical time-projection chamber, a single-arm electromagnetic calorimeter, a time-of-flight detector, a transition radiation detector, a single-arm ring imaging Cherenkov counter and a forward muon arm.

After the LHC startup and well before the design instantaneous luminosity of $10^{34} cm^{-2} s^{-1}$ is achieved, there are several physics channels with the potential for discovery. New physics producing resonances could include a new $Z^\prime$, \eg any new heavy gauge boson or from models with compact extra dimensions. Spectacular multi-jet, multi-lepton and missing energy signatures would also be produced by SUSY, with production of ~100 events per day at luminosities of $10^{33} cm^{-2} s^{-1}$ for squark and gluino masses of ~ 1 TeV. Almost all of the allowed Higgs mass range will be explored by ATLAS and CMS with 10 $\fb^{-1}$, and with 30 $\fb^{-1}$ is covered to more than 7 $\sigma$ over the whole range. The LHCb experiment will use the large sample of b-hadrons produced at the LHC to make precision studies of CP asymmetries and rare decays of B-mesons.  The ALICE experiment will search for new physics in strongly interacting matter at extreme energy densities where quark-gluon plasma is expected.

The new physics discoveries of the LHC will depend on the calibration and alignment of the detectors, the validation of  QCD and other Standard Model (SM) Monte Carlos (MC) and calculations and the overall understanding of the LHC detector and machine environment. A primary goal of the initial running will be to establish these items. This paper examines this process for the ALICE, ATLAS, CMS and LHCb experiments and looks a the measurements that can be anticipated from data sets taken with integrated luminosities of 1, 10 and 100 $\pb^{-1}$. 

\section{LHC Luminosity Profile}
\label{sec:lumi}

The expected profile for the phases of LHC operation from startup to design luminosity, assuming $6 \times 10^6$ seconds of pp operation per year is found in Table \ref{tab:lumibuild}. Since the focus of this article is the initial startup, a breakdown of the various steps taken during the first phase of LHC operation,  ``Phase A'' is shown in Table \ref{tab:lumistart}. It is expected to take at minimum about 30 days of beam time to establish collisions followed by about one week for each of the steps shown. The luminosities shown should be scaled by 0.71 at 10 TeV with respect to 14 TeV due to the increased beam emittance at the lower energy. To exploit the early part of the LHC startup, the experiment collaborations are examining what can be done by with a small integrated luminosity. Table \ref{tab:onepb} shows the expected yield of events from an integrated luminosity of 1 $\pb^{-1}$ at a center of mass energy of 14 TeV. The yields should be about 5\% lower for minimum bias events and a third lower for W and Z production at the expected initial collision energy of 10 TeV\cite{stirling}.

\begin{table}
\begin{center}
\begin{tabular}{|c|c|c|c|c|}
\hline
Parameter & Phase A & Phase B & Phase C & Nominal\\\hline
no. bunches & 43-156 & 936 & 2808 & 2808\\
Bunch Spacing & 2021-566 & 75 & 25 & 25 \\
N  $10^{11}$ protons) & 0.4-0.9 & 0.4 - 0.9 & 0.5 & 1.15\\
Luminosity ($cm^{-2} s^{-1}$) & $ 6 \times 10^{30} - 10^{32}$ & $10^{32} - 10^{33}$ & $ (1-2) \times 10^{33}$ & $10^{34}$\\
$ \int  \Lumi \delta t $ ($6 \times 10^6$ sec/year)  & $ < 100 \pb^{-1}/yr $ & $< 1 \fb^{-1}/yr$   & $< 10 \fb^{-1}/yr$  &  60 $\fb^{-1}/yr$\\\hline
\end{tabular}
\caption{Variations of LHC parameters through the various phases until nominal design luminosity is reached\cite{Bailey}.}
  \label{tab:lumibuild}
\end{center}
\end{table}

\begin{table}
\begin{center}
\begin{tabular}{|c|c|c|c|c|c|c|}
\hline
Bunches & $ \beta^*$ & $I_p$ & Lumi & Pileup & MB Rate & $ \int  \Lumi \delta t/wk$\\\hline
$ 1 \times 1$ & 18 & $10^{10}$ & $10^{27}$ &Low & 55 Hz & \\
$43 \times 43$ & 18 & $3 \times 10^{10}$  & $3.8 \times 10^{29}$ & 0.06 & 20 kHz & \\
$43 \times 43$ & 4 & $3 \times 10^{10}$  & $1.7 \times 10^{30}$ & 0.28 & 60 kHz &  $\sim 0.1\pb^{-1}$ \\
$43 \times 43$ & 2 & $4 \times 10^{10}$  & $6.1 \times 10^{30}$ & 0.99 & 200 kHz &  \\
$156 \times 156$ & 4 & $4 \times 10^{10}$  & $1.1 \times 10^{31}$ & 0.50 & 400 kHz &  $\sim 1\pb^{-1}$\\
$156 \times 156$ & 4 & $9 \times 10^{10}$  & $5.6 \times 10^{31}$ & 2.3 & 2 MHz & \\
$156 \times 156$ & 2 & $9 \times 10^{10}$  & $1.1 \times 10^{32}$ & 5.0 & 4 MHz &  $\sim 10\pb^{-1}$\\\hline
\end{tabular}
\caption{Variations of LHC parameters through the initial startup ``Phase A''\cite{Bailey}.}
 \label{tab:lumistart}
\end{center}
\end{table}

\begin{table}
\begin{center}
\begin{tabular}{|c|c|}
\hline
Channel & Number of Events\\\hline
W $\rightarrow \mu\nu$ & 7000 \\
Z $\rightarrow \mu\mu$ & 1100 \\
$t\bar{t} \rightarrow \mu + X$ & 80 \\
QCD Jets ($p_T > 150$ GeV) & 1000 (for 10\% of trigger bandwidth) \\
Minimum Bias & Trigger Limited \\
gluino-gluino (M $\sim$ 1 TeV) & 1-10 \\\hline
\end{tabular}
\caption{Expected yields of events from 1 $\pb^{-1}$ at a center of mass energy of 14 TeV, based on the ATLAS trigger and adapted from \cite{hepph0504221}.} 
 \label{tab:onepb}
\end{center}
\end{table}

\section{ATLAS and CMS Analysis of 1 and 10  $\pb^{-1}$ samples}
\label{sec:10pb}

The initial sample of events from the first 1 $ \pb^{-1}$ will be of considerable value to the LHC experiments. This early data can be used to improve the inter-calibration of the electromagnetic and hadronic calorimeters by looking at the azimuthal symmetry of deposited energy with minimum bias events. The calibration of the electromagnetic calorimeter can be improved by first using $\pi^0$ data and then electrons from $Z \rightarrow ee$. The jet energy scale can be improved using di-jet balancing, $\gamma/Z + jet$ events and  $W \rightarrow jj$ in $t\bar{t}$ events. The inner tracking and muon system alignment can be improved using $Z \rightarrow \mu\mu$ events. A summary of the calibration and alignment data and quantities improved by this data is listed in Table~\ref{tab:aligncalib}.

\begin{table}
\begin{center}
\begin{tabular}{|c|c|c|c|}
\hline
Measurement & ATLAS & CMS & Physics Samples\\\hline
ECAL Uniform. & 1-2\% & 4\% & Isolated electrons, $Z \rightarrow ee$ \\
$e/\gamma$ E-scale & $\sim$ 2\% & $\sim$ 2\% &   $Z \rightarrow ee$ \\
HCAL Uniform. & $\sim$ 3\% & $\sim$ 2-3\% &  single pions, QCD Jets \\
Jet E-scale & $<$ 10\% & $<$ 10\% & $\gamma/Z + 1j, W \rightarrow$ jj in $ t\bar{t}$ events \\
Track Align & 10-200 $\mu$m in R$\phi$ & 20-200 $\mu$m in R$\phi$ & Gen. Tracks, iso. $\mu, Z \rightarrow \mu\mu $ \\\hline
\end{tabular}
\caption{Measured quantities, expected knowledge of these quantities for ATLAS and CMS at startup, and examples of the initial data samples that are expected to improve these measurements\cite{Fabiola}.} 
  \label{tab:aligncalib}
\end{center}
\end{table}

Once the basic calibration is addressed and measurements are possible, one of the first measurements to be made will be the spectra of charged hadrons. This will be very useful for adjusting the tunable parameters in the Monte Carlos. These spectra have never been explored for $\sqrt{s} > $ 2 TeV. They represent an important tool for calibration and understanding of the detector response. The data samples for these measurements will be taken with zero-bias or minimum bias triggers. Letting these triggers operate at a few Hz for about a month should provide enough statistics to provide significant adjustments of the LHC Monte Carlo programs, even when a low operation duty factor is included.

Another basic initial measurement is that of the underlying event activity. The amount of underlying event activity is difficult to extrapolate from the Tevatron measurements and therefore is best measured during LHC startup. Both ATLAS and CMS will measure $d^2N/d\eta d\phi$ and $d^2p_T^{sum}/d\eta d\phi$ in the region perpendicular to the jet activity, \eg between the leading jet ($|\Delta\phi| < 60^\circ$) and the away regions ($|\Delta\phi| > 120^\circ$). This is an important ingredient for jet and lepton isolation, energy flow and jet-tagging calculations. It provides a sensitive test of the multiple-parton interaction (MPI) tunes applied to QCD calculations.

\begin{figure}
\begin{center}
\begin{minipage}[c]{0.48\textwidth}
\psfig{figure=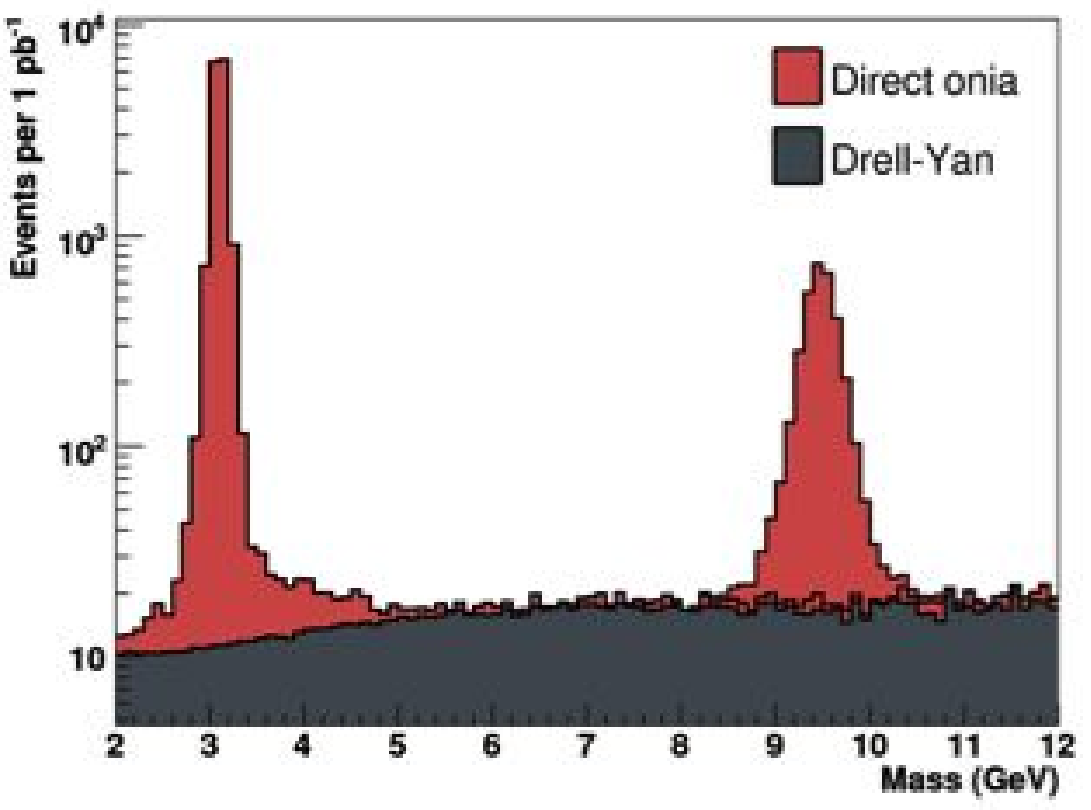,width={0.98\textwidth}}
\caption{$J/\psi$ and $\Upsilon$ peaks in invariant mass seen in the ATLAS detector after a simulated 1 $ pb^{-1}$ exposure\cite{Fabiola}.}
\label{fig:onia}
\end{minipage}
\hfill
\begin{minipage}[c]{0.48\textwidth}
\psfig{figure=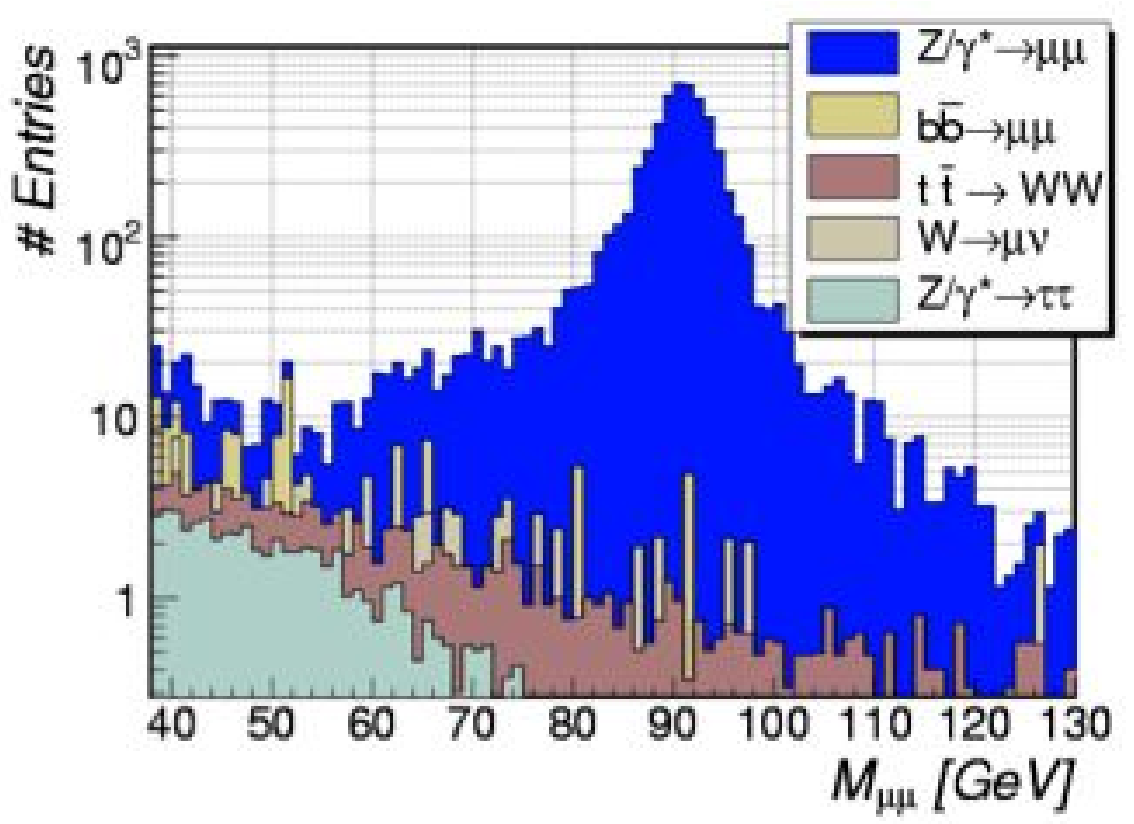,width={0.98\textwidth}}
\caption{Invariant dimuon mass spectrum for various SM peaks in the ATLAS detector after all cuts for an integrated luminosity of 10 $ pb^{-1}$\cite{Fabiola}.}
\label{fig:peaks}
\end{minipage}
\end{center}
\end{figure}

Standard Model onia peaks can be measured with 1 $ pb^{-1}$ of data. Figure~\ref{fig:onia} shows $J/\psi$ and $\Upsilon$ peaks in invariant mass seen in the ATLAS detector after a simulated 1 $ pb^{-1}$ exposure\cite{Fabiola}. After all cuts there are 15,600 (3100) $J/\psi$  ($\Upsilon$) events accumulated. There will also be about 600 $ z \rightarrow \mu\mu $ events that can be used for the alignment of the muon spectrometer, studies of ECAL uniformity, the energy and momentum scale of the full detector, as well as lepton trigger and reconstruction  efficiency.  

\begin{figure}
\begin{center}
\begin{minipage}[c]{0.48\textwidth}
\psfig{figure=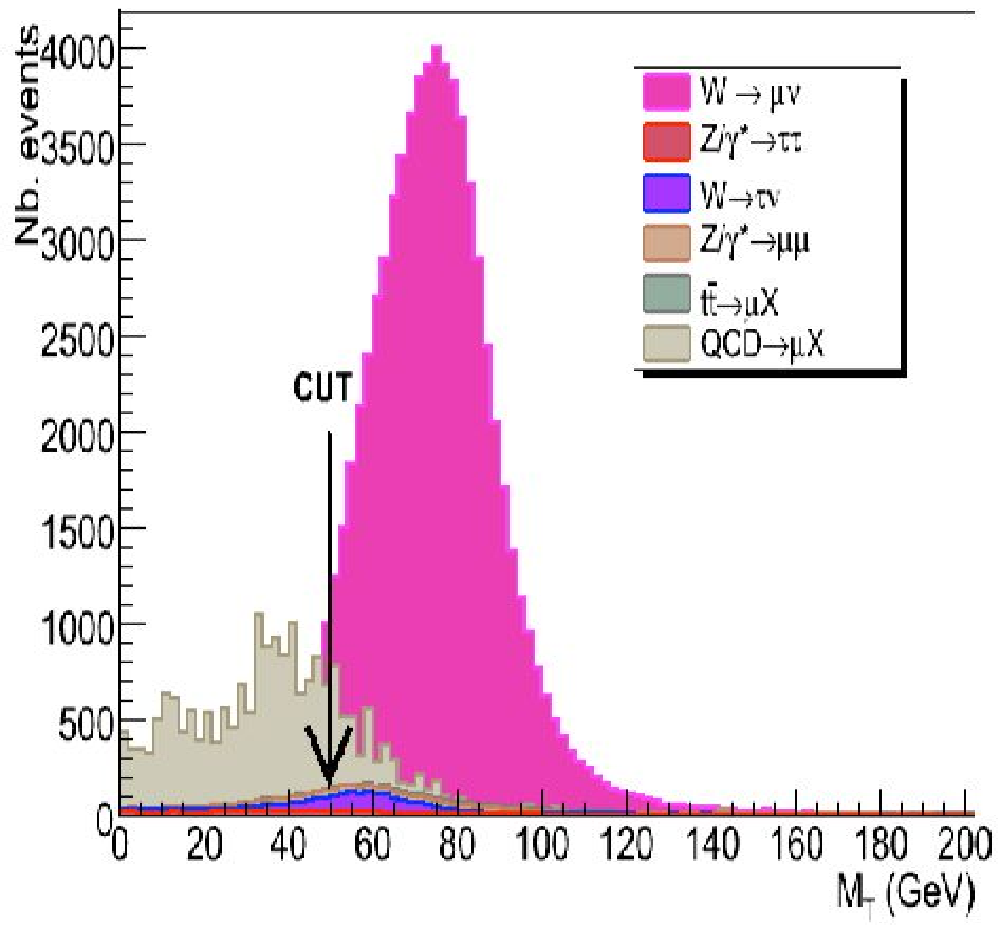,width={0.98\textwidth}}
\caption{Simulated $M_T$ distribution in the $W \rightarrow \mu\nu$ channel in CMS with an integrated luminosity of 10 $ pb^{-1}$\cite{PAS0701}.}
\label{fig:zmunu}
\end{minipage}
\hfill
\begin{minipage}[c]{0.48\textwidth}
\psfig{figure=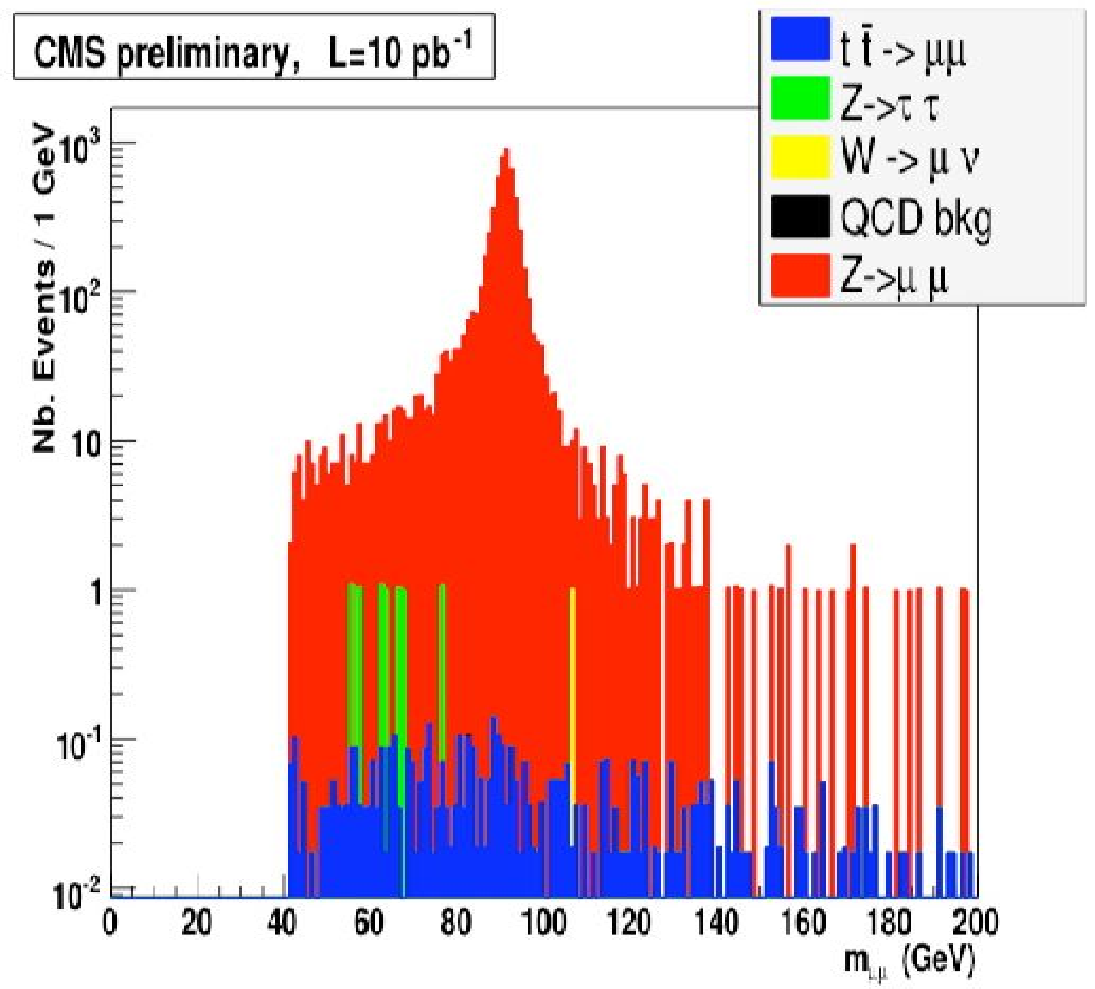,width={0.98\textwidth}}
\caption{Simulated di-electron invariant mass distribution at the Z-peak expected in CMS with an integrated luminosity of 10 $ pb^{-1}$\cite{PAS0702}.}
\label{fig:zee}
\end{minipage}
\end{center}
\end{figure}

The number of SM peaks observable increases substantially when an integrated luminosity of 10 $ pb^{-1}$ is accumulated. Figure~\ref{fig:peaks} shows the invariant dimuon mass spectrum for various SM peaks in the ATLAS detector after all cuts for an integrated luminosity of 10 $ pb^{-1}$\cite{Fabiola}. Figure~\ref{fig:zmunu} shows the simulated transverse mass distribution in the  $W \rightarrow \mu\nu$ channel  expected by CMS with an integrated luminosity of 10 $ pb^{-1}$\cite{PAS0701}. Using this luminosity, ATLAS and CMS can measure the W and Z cross sections with sufficient statistics that the uncertainty will be dominated by the luminosity measurement (10\%). The muon channels should be very clean and their largest measurement uncertainty should be the momentum scale error estimated at less than (2.7\%). 

The electron channels should benefit from selection algorithms that are robust with respect to miscalibrations and misalignments. However, the QCD backgrounds are significant for the W signals. Figure~\ref{fig:zee} shows the simulated di-electron invariant mass distribution at the Z-peak expected by CMS with an integrated luminosity of 10 $ pb^{-1}$\cite{PAS0702}. With higher luminosities these measurements are useful for determining Parton Distribution Functions (PDFs).

\begin{figure}
\begin{center}
\begin{minipage}[c]{0.48\textwidth}
\psfig{figure=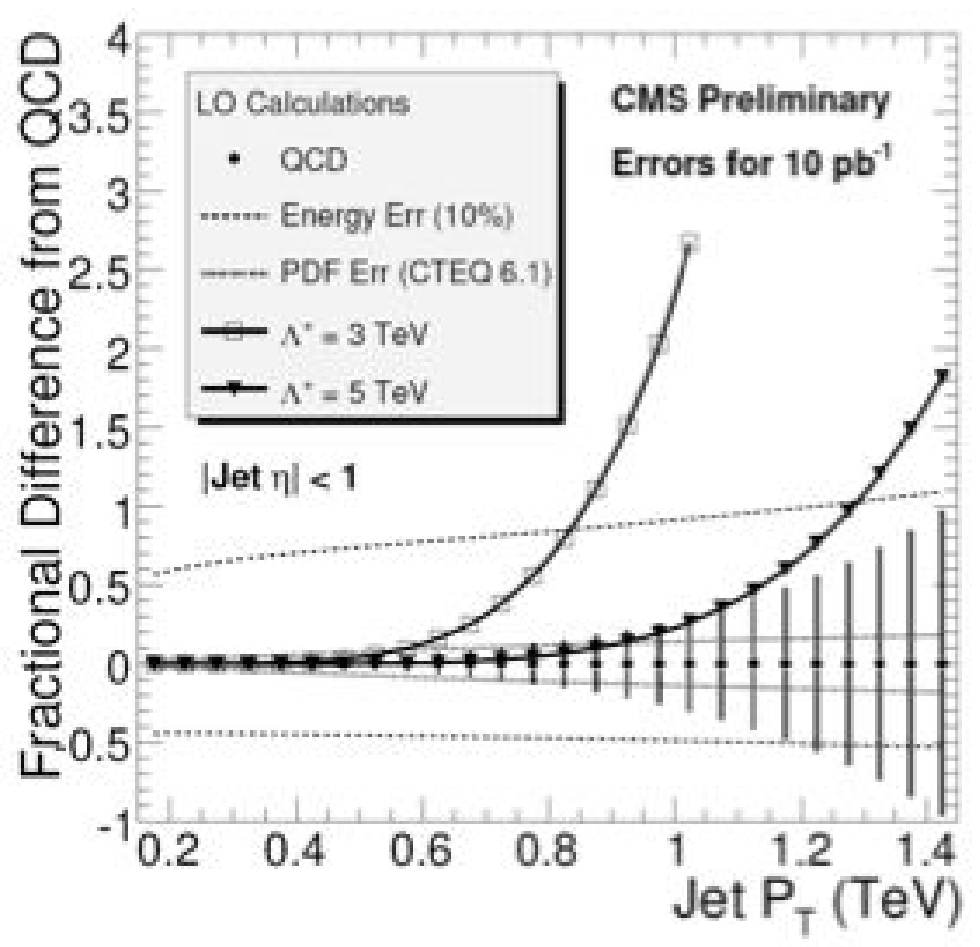,width={0.98\textwidth}}
\caption{Simulated fractional difference from the QCD expectation with PDF and Systematic (JES) errors plotted against Jet $p_T$ for $\Lambda^+$ = 3 and 5 TeV for an integrated luminosity of 10 $ pb^{-1}$ in CMS\cite{CMS08019}.}
\label{fig:contact}
\end{minipage}
\hfill
\begin{minipage}[c]{0.48\textwidth}
\psfig{figure=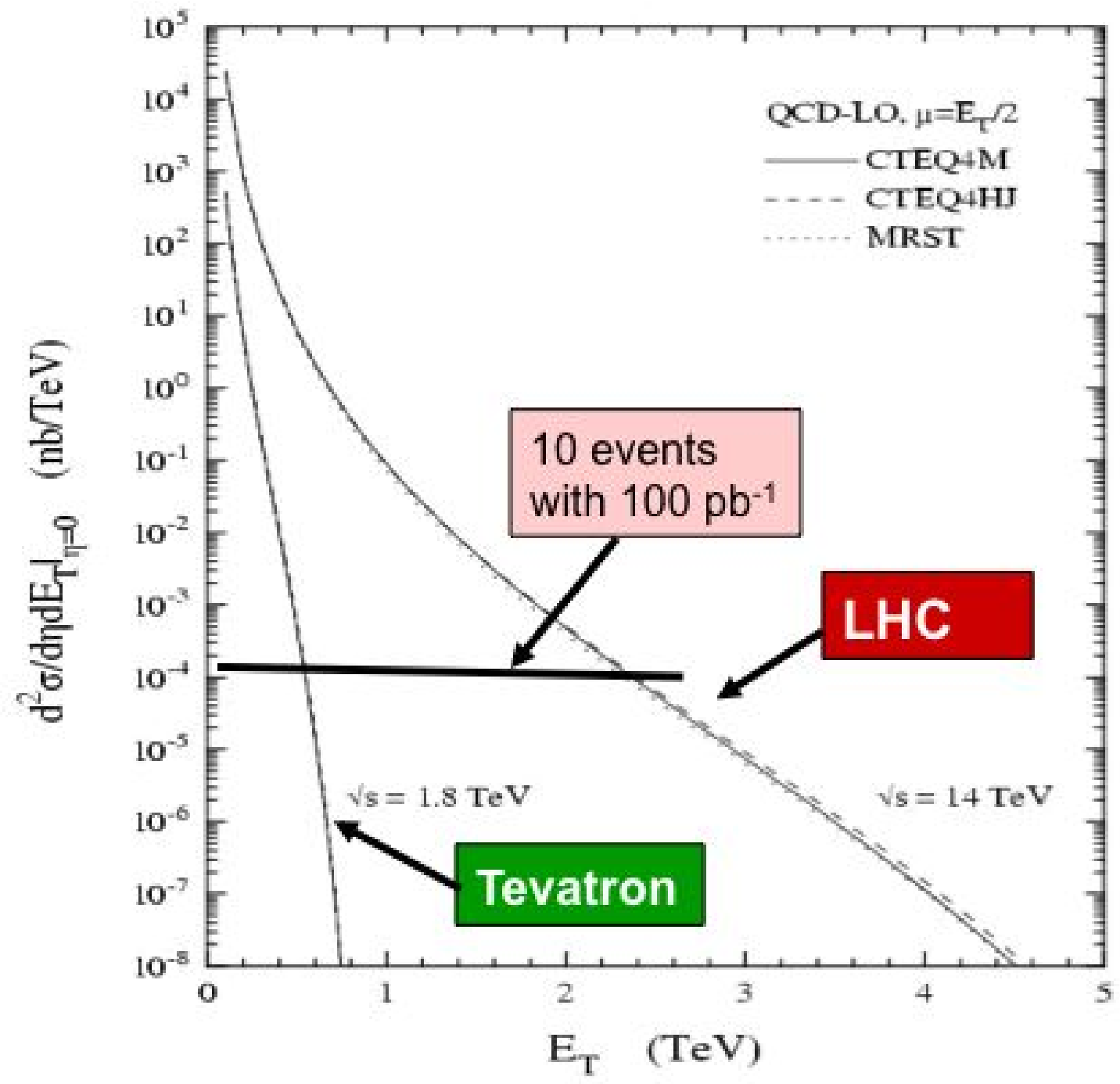,width={0.98\textwidth}}
\caption{Calculated jet cross sections vs. $E_T$ for the Tevatron and  LHC experiments using 3 different PDFs. A line indicates where 10 events are expected for an an integrated luminosity of 100 $ pb^{-1}$\cite{Fabiola}.}
\label{fig:jetxsec}
\end{minipage}
\end{center}
\end{figure}

Contact interactions create a large rate at high $p_T$, making a discovery possible with an integrated luminosity of 10 $ pb^{-1}$. The error in the observation is dominated by the uncertainty in the Jet Energy Scale (JES)  ($\sim$ 10\%) in this early running. The PDF uncertainties and statistical errors are smaller than the JES uncertainty.  With an integrated luminosity of 10 $ pb^{-1}$, new physics can be observed beyond the Tevatron exclusion of $\Lambda^+ <  2.7 $ TeV\cite{CMS08019}. Figure~\ref{fig:contact} shows the simulated fractional difference from the QCD expectation with PDF and systematic (JES) errors plotted against Jet $p_T$ for $\Lambda^+$ = 3 and 5 TeV for an integrated luminosity of 10 $ pb^{-1}$ in CMS.

\section{ATLAS and CMS Analysis of 100  $\pb^{-1}$ Samples}
\label{sec:100pb}

The ATLAS and CMS collaborations will use the data sample from an integrated  luminosity of 100 $ pb^{-1}$ to commission and calibrate their detectors in situ using well-known physics samples. This process will be time-consuming, but is essential preparation for searching for discovery physics. The $Z \rightarrow ee, \mu\mu$ samples will be useful for the tracking, electromagnetic calorimetry and muon chamber calibration and alignment. Events where $t\bar{t} \rightarrow bl\nu bjj$ will help to set the JES from $W \rightarrow jj$ and to measure the b-tagging performance. The experiments will also ``rediscover'' and measure SM physics at 10 and then 14 TeV, such as W, Z, $t\bar{t}$, and QCD jets, particularly since these are backgrounds to new physics.  A summary of calculated samples of events written to tape by ATLAS for an integrated  luminosity of 100 $ pb^{-1}$ is shown in Table~\ref{tab:100pb}\cite{Fabiola}.

\begin{table}
\begin{center}
\begin{tabular}{|c|c|c|}
\hline
Channels (examples) & Events to tape & Total LEP \& Tevatron Statistics\\\hline
$W \rightarrow \mu\nu $ & $\sim 10^6$  &$\sim 10^4$ LEP, $ \sim 10^{6-7}$ Tevatron \\
$Z \rightarrow \mu\mu $ & $\sim 10^5$  &$\sim 10^6$ LEP, $ \sim 10^{5-6}$ Tevatron \\
$t\bar{t} \rightarrow W b W b \rightarrow \mu\nu + X $ & $\sim 10^4$  &$\sim 10^4$ LEP, $ \sim 10^{6-7}$ Tevatron \\
$\tilde{g}\tilde{g}$,  m($\tilde{g}$)$\sim$1~TeV & 50 & \\\hline
\end{tabular}
\caption{Samples of events to tape for ATLAS for an an integrated  luminosity of 100 $ pb^{-1}$\cite{Fabiola}.} 
  \label{tab:100pb}
\end{center}
\end{table}

\begin{figure}
\begin{center}
\begin{minipage}[c]{0.48\textwidth}
\psfig{figure=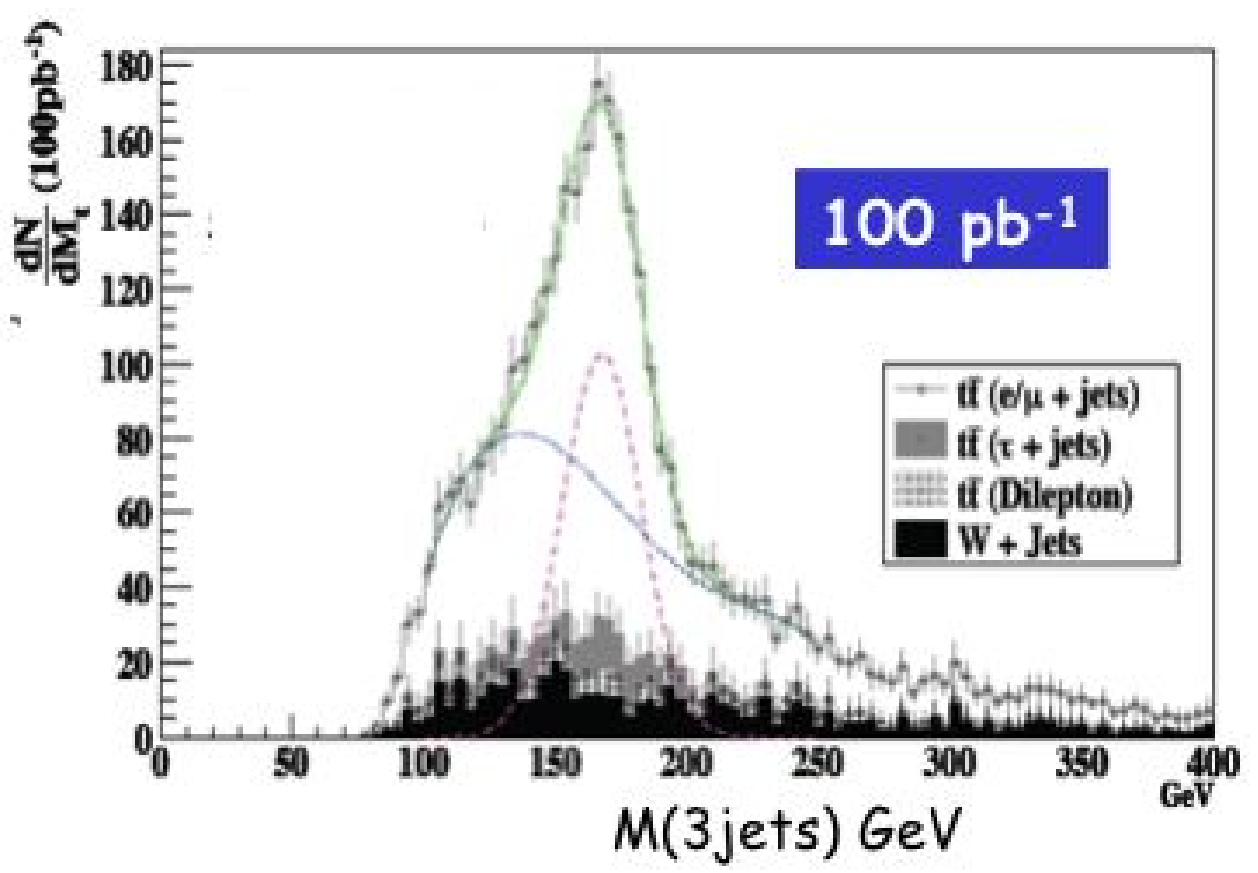,width={0.98\textwidth}}
\caption{Simulated invariant mass distribution of the three jets with the highest $\Sigma p_T$ for various top signals and W + jets from an integrated luminosity of 100 $ pb^{-1}$ in the ATLAS detector\cite{Fabiola}.}
\label{fig:top}
\end{minipage}
\hfill
\begin{minipage}[c]{0.48\textwidth}
\psfig{figure=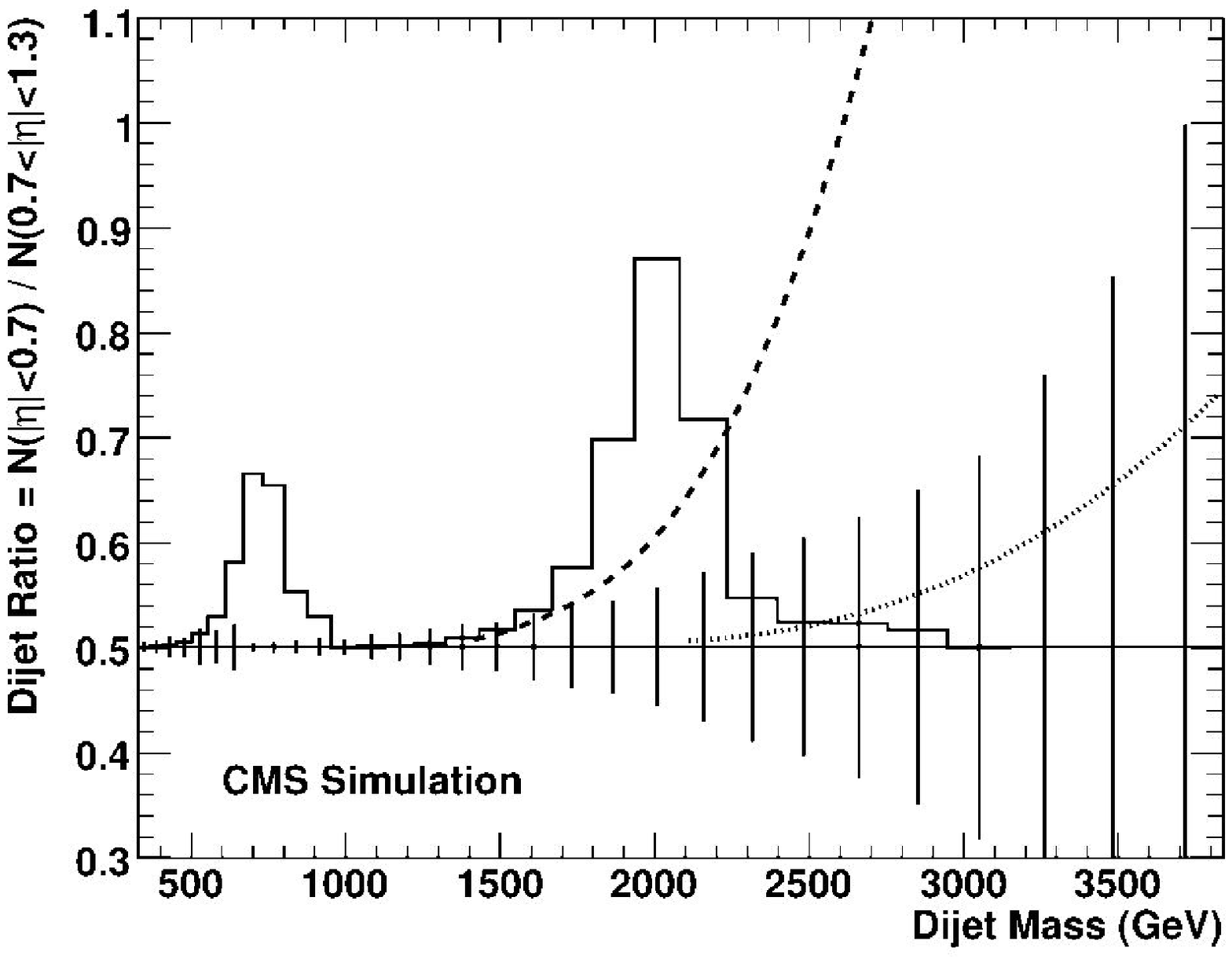,width={0.98\textwidth}}
\caption{Simulated CMS dijet ratio for corrected QCD jets (horizontal line), with statistical uncertainties (vertical bars) for an integrated luminosity of 100 $ pb^{-1}$, compared to QCD + contact interaction signals with a scale $\Lambda^+$ = 5 TeV (dashed) and 10 TeV (dotted), as well as to QCD + dijet resonance signals (histogram) with q* masses of 0.7 and 2 TeV \cite{CMS08019}.}
\label{fig:dijetres}
\end{minipage}
\end{center}
\end{figure}

Top events are central and back to back so they can be observed quickly even with limited detector performance and analysis machinery. Using cuts such as 3 jets with $p_T >  40$ GeV, 1 jet with $p_T >  20$ GeV, an isolated lepton with $p_T >  20$ GeV, and missing energy, $E_T^{miss} > 20$ GeV can isolate a viable signal of $\sim 3000$ events without a b-tag using an integrated  luminosity of 100 $ pb^{-1}$\cite{Fabiola}.  Figure~\ref{fig:top} shows the simulated invariant mass distribution of the three jets with the highest $\Sigma p_T$ for various top signals and W + jets from an integrated luminosity of 100 $ pb^{-1}$ in the ATLAS detector\cite{Fabiola}.  This should enable a 20\% measurement of $\sigma_{t\bar{t}}$, and a measurement of $m_t$ to $< 10$ GeV. This will help to constrain the theory and MC generators using such observables as the $p_T$ spectra.

This early LHC data can also be used to constrain the Parton Distribution Functions using the angular distributions of $W \rightarrow l\nu$ events. Since $x_{1,2}  = \frac{M}{{\sqrt S }}e^{ \pm y}$, W production over $|y| < 2.5$ at the LHC  involves $10^{-4} < x_{1,2} <  0.1$, which is the region dominated by $g \rightarrow q\bar{q}$. The PDF uncertainties in this region are in the 4 - 8 \% range. Therefore early measurements of $e^\pm$ angular distributions at the LHC can provide discrimination between different PDFs if the experimental precision is $\sim 4\%$. The systematics of these events (\eg $e^\pm$ acceptance vs.$\eta$) should be controllable to a few percent with $Z \rightarrow ee$ events ($\sim30,000$ events expected in 100 $pb^{-1}$)\cite{Fabiola}.

It is important to note that with 100 $ pb^{-1}$ the ATLAS and CMS experiments are exploring new territory. Figure~\ref{fig:jetxsec} shows the calculated jet cross sections in $E_T$ for the Tevatron and  LHC experiments with 3 different PDFs\cite{Fabiola}. A line indicates where 10 events are expected for an integrated luminosity of 100 $ pb^{-1}$. The dijet cross sections can be examined for resonances. Strongly produced resonances can be seen with a convincing signal for an excited quark at a mass of up to 2 TeV with 100  $ pb^{-1}$ of data. Figure~\ref{fig:dijetres} shows the CMS simulation of dijet resonances reconstructed using corrected jets, coming from q* signals of mass 0.7 and 2 TeV. The dijet ratio can be optimized for a contact interaction search by concentrating on the barrel region since the QCD cross section rises dramatically with the $|\eta|$ cut due to the t-channel pole \cite{CMS08019}. CMS has shown that the sensitivity to a contact interaction signal is enhanced by this with the value of a discoverable $\Lambda^+$ increasing to 7 TeV. Figure~\ref{fig:dijetres} shows the CMS simulation of the dijet ratio for corrected jets expected from QCD compared to QCD + contact interaction signals with a scale $\Lambda^+$ = 5 TeV and 10 TeV.

\begin{figure}
\begin{center}
\begin{minipage}[c]{0.48\textwidth}
\psfig{figure=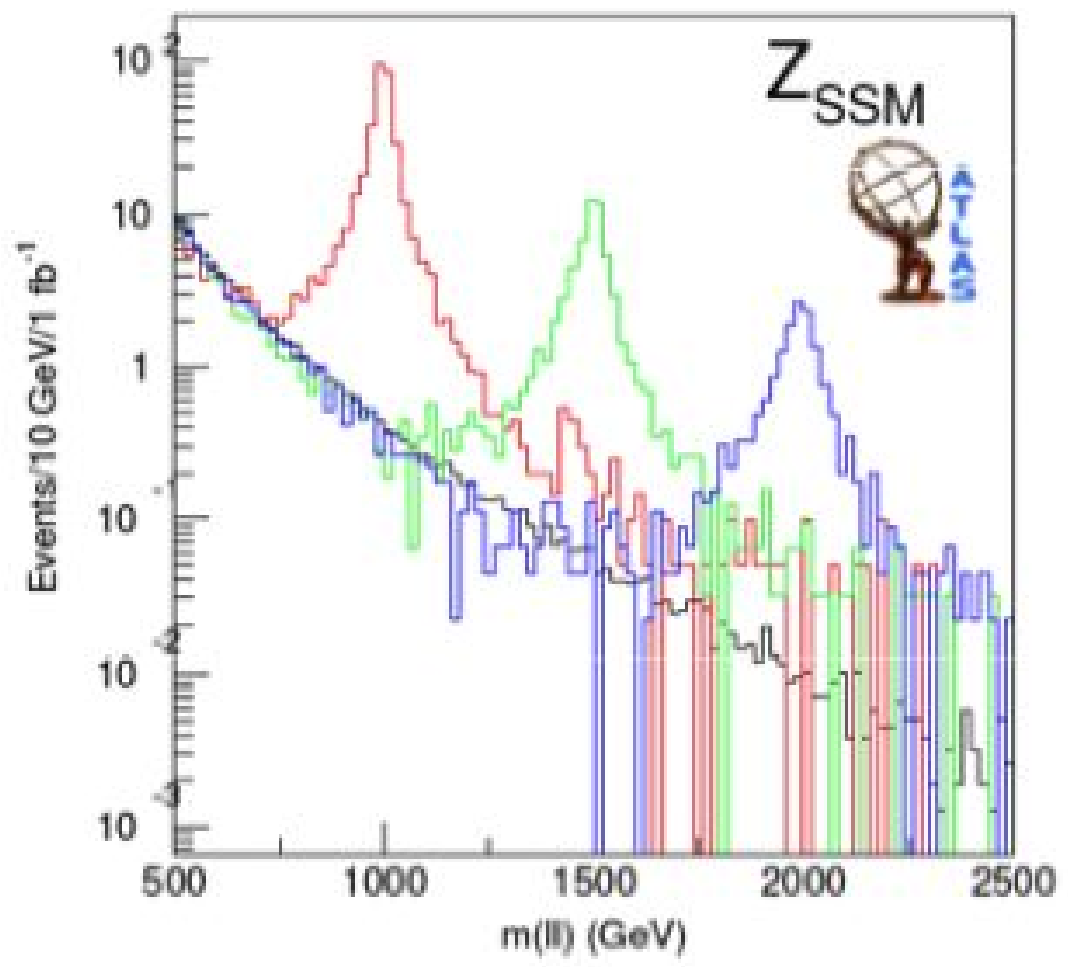,width={0.98\textwidth}}
\caption{Simulated number of dilepton events produced in ATLAS by a resonance at 1, 1.5 and 2 TeV from an integrated luminosity of 100 $ pb^{-1}$ \cite{Fabiola}.}
\label{fig:resonances}
\end{minipage}
\hfill
\begin{minipage}[c]{0.48\textwidth}
\psfig{figure=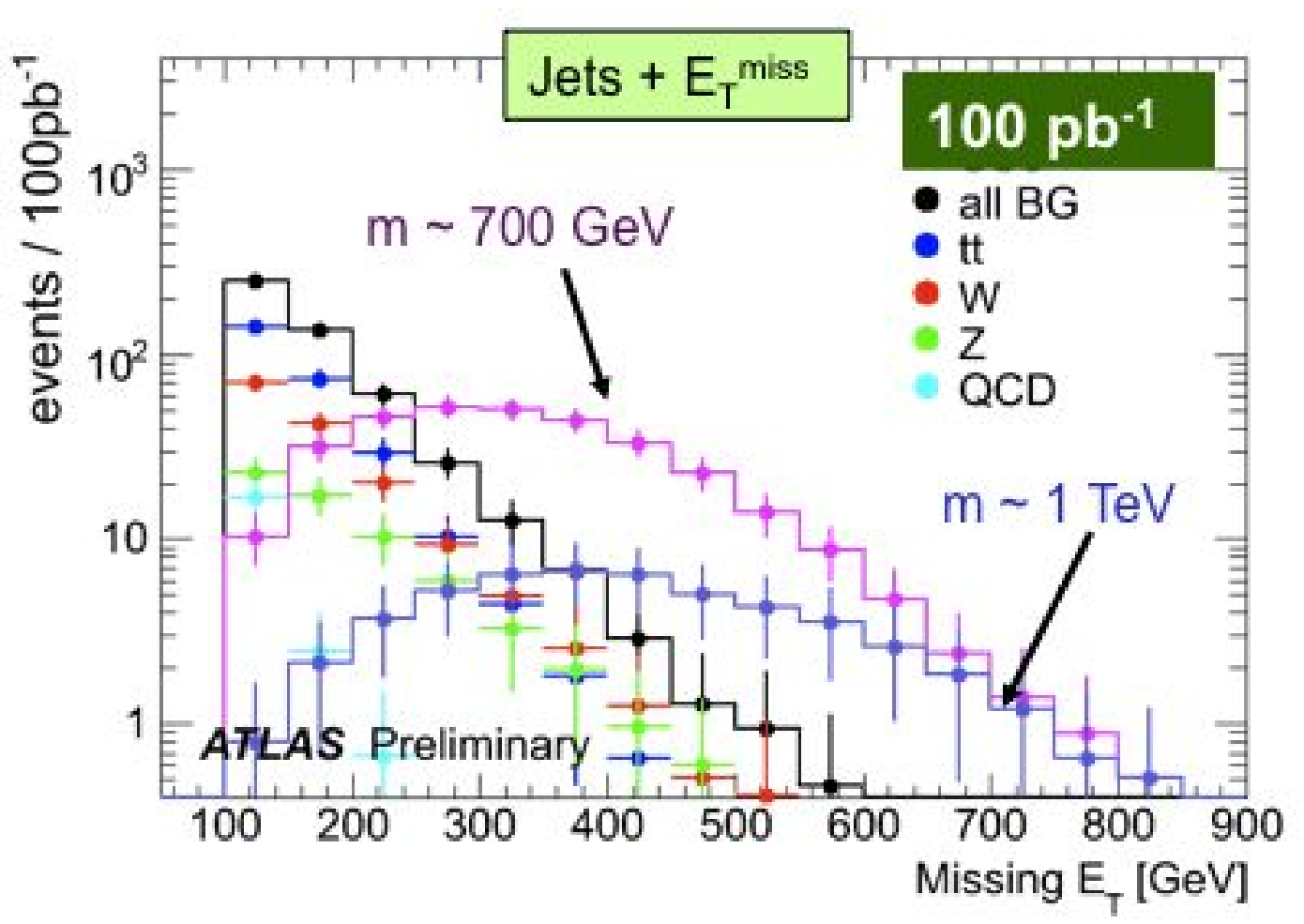,width={0.98\textwidth}}
\caption{Missing $E_T$ distribution for simulated SUSY signals and backgrounds in events with jets and missing $E_T$ in ATLAS for an integrated luminosity of 100 $ pb^{-1}$ \cite{Fabiola}.}
\label{fig:susy}
\end{minipage}
\end{center}
\end{figure}

New physics such as a narrow resonance decaying into $e^+e^-$ should be discoverable up to a mass of 1 TeV. The signal is a narrow mass peak on top of a small Drell-Yan background. Figure~\ref{fig:resonances} shows the number of dilepton events produced in ATLAS by a resonance at 1, 1.5 and 2 TeV from an integrated luminosity of 100 $ pb^{-1}$. Further understanding of the source of the resonance, (\eg a $Z^\prime$ or a graviton) involves use of the angular distribution and a much larger data sample (100 $fb^{-1}$).

If present, SUSY should produce a large cross section for $ \tilde{q}\tilde{q}, \tilde{g}\tilde{q}$ and $\tilde{g}\tilde{g}$ production. These events would have spectacular signatures with many jets, leptons and missing $E_T$. Hints of SUSY production would begin to manifest themselves up to $m \sim$ 1 TeV for an integrated luminosity of 100 $ pb^{-1}$ as shown in Figure~\ref{fig:susy}. However, a luminosity of $\sim$ $1 fb^{-1}$ is needed to properly understand the backgrounds\cite{Fabiola}.

\section{Early Physics with LHCb}
\label{LHCb}

The LHCb detector will be able to record $10^8$ events in $\sim 20$ hours at a luminosity of $10^{28}cm^{-2} s^{-1}$ with its interaction trigger. This would provide a first look at minimum bias events in 10 TeV (and later 14 TeV) collisions. The LHCb collaboration will measure the ratios of charged track multiplicities ($\pi/K/p, +/-$) vs. $\eta, p_T, \varphi$. They will perform reconstruction and production studies of $K_S, \Lambda, \phi, D$ and other particles. These measurements will provide valuable input to the theoretical calculations and Monte Carlo generators that are essential to understanding the physics in this new kinematic regime. As an example, LHCb should have a sample of $\sim$ 1M $J/\psi \rightarrow \mu\mu$ events in 1 $ pb^{-1}$ of data. These will be used to measure the fraction of $J/\psi$ from b decays or prompt production vs. $p_T$, observe the first exclusive $B \rightarrow J/\psi X$ peaks and measure the $b\bar{b}$ production cross section. Figure~\ref{fig:LHCb} shows the extraction of the $J/\psi$ peak from  a full simulation of 12.8M minimum bias events in LHCb. This shows that even with a small amount of integrated luminosity, a significant peak can be extracted.

One important signal for LHCb is $B_S \rightarrow \mu\mu$. The efficiency for these events of 10\% is large so the principal experimental issue is background rejection. The largest background is $ b \rightarrow \mu, b \rightarrow \mu$ with the specific background of $B_c \rightarrow J/\psi(\mu\mu)\mu\nu$ the most dominant. LHCb will exploit its muon identification, vertexing and mass resolution of 18 $Mev/c^2$ to extract a signal that is anticipated to exceed the combined Tevatron experiment statistics after an integrated luminosity of 50 $ pb^{-1}$\cite{Schneider}.

\section{Early Physics with ALICE}
\label{sec:ALICE}

\begin{figure}
\begin{center}
\begin{minipage}[c]{0.48\textwidth}
\psfig{figure=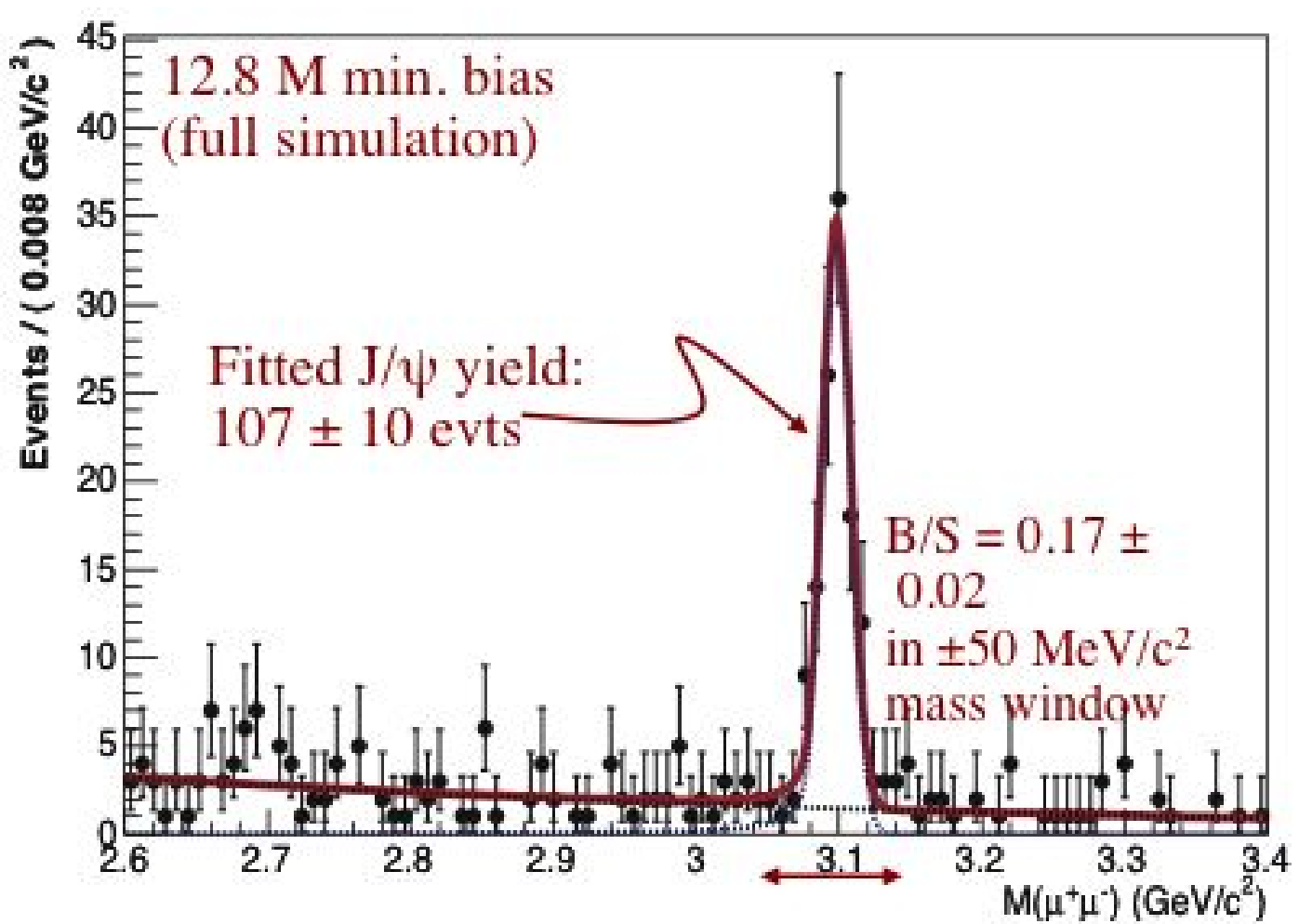,width={0.98\textwidth}}
\caption{extraction of the $J/\psi$ peak from  a full simulation of 12.8M minimum bias events in LHCb. \cite{Schneider}.}
\label{fig:LHCb}
\end{minipage}
\hfill
\begin{minipage}[c]{0.48\textwidth}
\psfig{figure=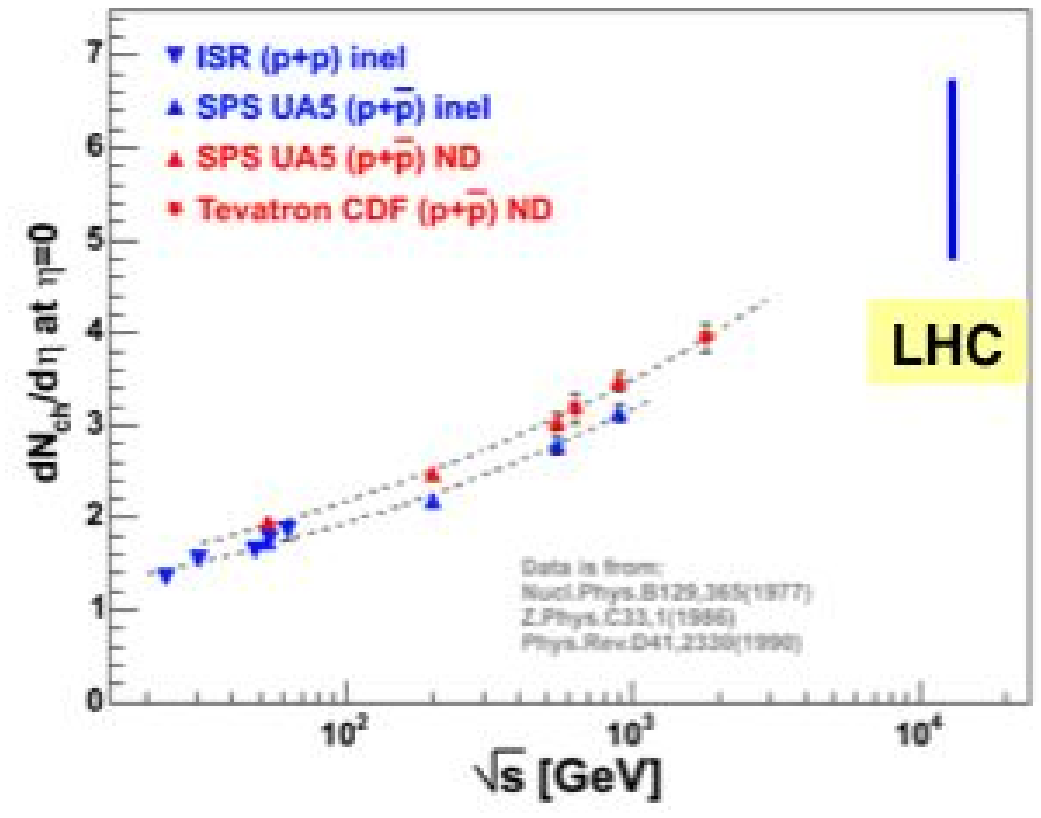,width={0.98\textwidth}}
\caption{Extrapolation of existing charged multiplicity measurements to the LHC \cite{Wessels}.}
\label{fig:ALICE}
\end{minipage}
\end{center}
\end{figure}

The first physics with ALICE will use pp collisions. This will provide important ``reference" data for the heavy ion program to follow. ALICE has extensive particle ID capabilities that will provide measurements for tuning minimum bias Monte Carlos and jet fragmentation models. ALICE can separate $\pi^\pm$, K and p between 100 MeV and 50 GeV. The transition radiation detector can distinguish electrons above 1 GeV and muons above 5 GeV. Weak decays and resonances can be identified between 0.5 to above 10 GeV and $\pi^0$s can be detected between 1 and 80 GeV\cite{Wessels}. This provides a unique capability for ALICE to measure multiplicity distributions, baryon transport and the charm cross section that is a major input to pp QCD calculations. The measurement of charged multiplicity will be extended a factor of 7 in $\sqrt{s}$. Figure~\ref{fig:ALICE} shows the extrapolation of existing charged multiplicity measurements to the LHC. This will extend the existing energy dependence measurements and provide a new look at the particle fluctuations in pp collisions. 

Measurements of the Baryon-Antibaryon asymmetry in ALICE should be able to distinguish between baryon number transport models, such as those based on quark exchange vs. string junction exchange. There is a large rapidity gap at the LHC ($> 9$ units) measurable in ALICE. The predicted Baryon-Antibaryon asymmetry for string junction exchange is 3-7\% and it is predicted to be multiplicity dependent. ALICE has estimated the contributions to the systematic error in the asymmetry to be less than 1\% from uncertainties in the cross section, the detector material budget and beam gas events\cite{Wessels}. The statistical error is $< 1\%$ for $10^6$ pp events, which could be taken in one day. This measurement could also be extended to $\Lambda\bar{\Lambda}$ where the asymmetry is larger.

\section{Conclusions}

The early physics data at the LHC will be used by ALICE, ATLAS, CMS and LHCb experiments to lay the groundwork for understanding the physics of the new energy frontier. As such, samples from integrated luminosities of 1, 10, and 100 $ pb^{-1}$ will have great value. If we are exceedingly fortunate they may even provide a hint of the new physics awaiting us.

\section*{Acknowledgements}

I would like to thank Sridhara Dasu, Fabiola Gianotti, Joe Incandela, Karl Jakobs, Bruce Mellado,  Tatsuda Nakada, Olivier Schneider, Jurgen Schukraft, Paris Sphicas and Johannes Wessels for their assistance, insights, advice and plots in the preparation of this paper.

\end{document}